\documentclass[a4paper,11pt]{article}
\usepackage{aaskaiid}

\usepackage{xcolor}
\usepackage{enumitem}
\usepackage{bm}
\usepackage{hyperref}
\usepackage{subcaption}
\usepackage{comment}
\usepackage[utf8]{inputenc}
\usepackage{orcidlink}
\usepackage{newunicodechar}
\graphicspath{{Figures/}} 

\newunicodechar{́}{\'}

\title{Foreground Characterization and Mitigation in the Observations of the CD/EoR with the SKA}
\ShortTitle{Foreground characterization and mitigation for SKA}
\ShortName{Bull et al.} 

\author[1,2]{Philip Bull\orcidlink{0000-0001-5668-3101}}

\author[1]{Jacob Burba\orcidlink{0000-0002-8465-9341}}

\author[3]{Emilio Ceccotti\orcidlink{0000-0002-3351-5778}}

\author[4,5]{Arnab Chakraborty\orcidlink{0000-0002-7758-9859}}

\author[6]{Samir Choudhuri\orcidlink{0000-0002-2338-935X}}

\author[7]{Abhirup Datta\orcidlink{0000-0002-5333-1095}}
\emailAdd{abhirup.datta@iiti.ac.in}

\author[6]{Khandakar Md Asif Elahi\orcidlink{0000-0003-1206-8689}}

\author[8,9]{Sambit K. Giri\orcidlink{0000-0002-2560-536X}}

\author[10]{Takumi Ito\orcidlink{0009-0001-8913-1044}}

\author[11]{Vibor Jeli\'c\orcidlink{0000-0002-6034-8610}}

\author[12]{Yi Mao\orcidlink{0000-0002-1301-3893}}

\author[13,14]{Florent Mertens\orcidlink{0000-0003-3802-4289}}

\author[14]{Satyapan Munshi\orcidlink{0000-0001-9919-4121}}

\author[15,16]{Ridhima Nunhokee\orcidlink{0000-0002-5445-6586}}

\author[17]{André R. Offringa\orcidlink{0000-0002-5809-2783}}

\author[7]{Samit Kumar Pal\orcidlink{0000-0002-2271-4165}}

\author[7]{Rashmi Sagar}

\author[18]{Huanyuan Shan\orcidlink{0000-0001-8534-837X}}

\author[19,20]{Peter H. Sims\orcidlink{0000-0002-2871-0413}}

\author[21]{Emma Tolley\orcidlink{0000-0002-1027-1213}}

\author[7]{Anshuman Tripathi\orcidlink{0000-0002-5091-9950}}

\author[22]{Le Zhang}

\author[18]{Zhenghao Zhu\orcidlink{0000-0001-8443-6095}}

\affiliation[1]{Jodrell Bank Centre for Astrophysics, University of Manchester, Manchester, M13 9PL, United Kingdom}

\affiliation[2]{Department of Physics and Astronomy, University of Western Cape, Cape Town 7535, South Africa}

\affiliation[3]{INAF -- Istituto di Radioastronomia, Via P. Gobetti 101, 40129 Bologna, Italy}

\affiliation[4]{Department of Physics, McGill University, Montreal, QC, Canada}

\affiliation[5]{Trottier Space Institute, McGill University, Montreal, QC, Canada}

\affiliation[6]{Centre for Strings, Gravitation and Cosmology, Department of Physics, Indian Institute of Technology Madras, Chennai 600036, India}

\affiliation[7]{Department of Astronomy, Astrophysics and Space Engineering, Indian Institute of Technology Indore, Indore 453552, India}

\affiliation[8]{Department of Astronomy and Oskar Klein Centre, AlbaNova, Stockholm University, SE-10691 Stockholm, Sweden}

\affiliation[9]{Van Swinderen Institute for Particle Physics and Gravity, University of Groningen, Nijenborgh 3, 9747 AG Groningen, The Netherlands}

\affiliation[10]{Kumamoto University, Graduate School of Science and Technology, 2-39-1 Kurokami, Chuo-ku, Kumamoto 860-8555, Japan}

\affiliation[11]{Ru{\dj}er Bo\v{s}kovi\'{c} Institute, Bijeni\v{c}ka cesta 54, 10000 Zagreb, Croatia}

\affiliation[12]{Department of Astronomy, Tsinghua University, Beijing 100084, China}

\affiliation[13]{LUX, Observatoire de Paris, PSL Research University, CNRS, Sorbonne Université, F-75014 Paris, France}

\affiliation[14]{Kapteyn Astronomical Institute, University of Groningen, PO Box 800, 9700 AV Groningen, The Netherlands}

\affiliation[15]{International Centre for Radio Astronomy Research, Curtin University, Bentley, WA, Australia}

\affiliation[16]{ARC Centre of Excellence for All Sky Astrophysics in 3 Dimensions (ASTRO 3D), Bentley, Australia}

\affiliation[17]{ASTRON, PO Box 2, 7990 AA Dwingeloo, The Netherlands}

\affiliation[18]{Shanghai Astronomical Observatory (SHAO), Nandan Road 80, Shanghai 200030, People's Republic of China}

\affiliation[19]{Cavendish Astrophysics, University of Cambridge, Cambridge, UK}

\affiliation[20]{Kavli Institute for Cosmology in Cambridge, University of Cambridge, Cambridge, UK}

\affiliation[21]{Institute of Physics, Laboratory of Astrophysics, École Polytechnique Fédérale de Lausanne (EPFL), Observatoire de Sauverny, Versoix, 1290, Switzerland}

\affiliation[22]{School of Physics and Astronomy, Sun Yat-sen University, Zhuhai 519082, P. R. China}

\abstract{The Square Kilometre Array (SKA), with its unprecedented sensitivity, frequency coverage, and large collecting area, is poised to revolutionize our understanding of the Cosmic Dawn (CD) and Epoch of Reionization (EoR) epochs marking the formation of the first luminous sources and the subsequent reionization of the intergalactic medium (IGM). However, detecting the faint redshifted 21 cm signal from neutral hydrogen remains one of the foremost challenges in observational cosmology, as it is buried beneath bright foregrounds from Galactic synchrotron radiation, free–free emission, and extragalactic point sources that are 4–5 orders of magnitude stronger than the cosmological signal. In this chapter, we highlight the key components and characteristics of these foregrounds and review ongoing efforts to model, characterize, and mitigate them. We emphasize how the SKA-Low AA* configuration, through its optimized array design, wide field of view, and improved calibration accuracy, enhances our capacity to suppress foreground contamination and recover the cosmological signal. The SKA Observatory’s Foreground Challenge plays a pivotal role in this effort by bringing together the global EoR/CD community to develop, compare, and validate foreground removal pipelines using realistic simulated datasets. Building on the experience of existing pathfinders such as LOFAR, MWA, and HERA, these collaborative initiatives are helping refine statistical and machine learning–based approaches for signal recovery. Together, these advancements are laying the groundwork for the SKA to probe the thermal and ionization history of the early Universe with unprecedented precision.}

\begin{document}
\maketitle
\section{Introduction}\label{sec:intro}
The Square Kilometre Array Observatory (SKAO) is set to revolutionize our understanding of the Universe, particularly the Cosmic Dawn (CD) and the Epoch of Reionization (EoR) two pivotal periods when the first stars, galaxies, and black holes emerged, transforming the thermal and ionization state of the intergalactic medium (IGM) \citep{2023Shaw, 2023Shimabukuro}. Conceived as the world’s largest and most sensitive radio telescope, the SKAO will offer unparalleled sensitivity, angular coverage, and spectral resolution, enabling detailed exploration of the redshifted 21-cm signal from neutral hydrogen (HI), a powerful probe of the early Universe \citep{2015Chapman, 2020Greig, 2023Rao}. This faint cosmological signal encodes the imprint of the first luminous structures and the evolution of the neutral IGM, making its detection one of the ultimate frontiers of modern observational cosmology \citep{2023Shaw, Mazumder2022}.

However, detecting the 21-cm signal is extremely challenging due to the dominance of bright astrophysical foregrounds, which exceed the cosmological signal by several orders of magnitude \citep{2015Chapman, Bonaldi2019}. These foregrounds originate primarily from Galactic synchrotron and free–free emission, along with contributions from extragalactic radio sources such as active galactic nuclei (AGN) and radio halos \citep{2019Li, 2004Jarvis}. In addition to these intrinsic sources, instrumental effects and calibration errors further complicate the recovery of the cosmological signal \citep{Datta2009, 2016MNRAS.461.3135B, Mazumder2022, Pal2025}. Even small calibration inaccuracies, on the order of $0.01\%$, can introduce significant residual systematics, leading to biased inference of EoR parameters \citep{Mazumder2022, 2025Tripathi, beohar2025}. Moreover, the widely used assumption of foreground spectral smoothness can break down in the presence of polarized emission affected by Faraday rotation, introducing additional spectral complexity \citep{2019Spinelli, 2018Sobey}.

Addressing these challenges requires the development of robust foreground mitigation and removal techniques capable of distinguishing the cosmological signal from both astrophysical and instrumental contaminants \citep{2015Chapman, Chen2023}. The SKA-Low array, designed specifically for low-frequency observations, will be instrumental in these efforts \citep{Acedo2017, Oscar2024, Oscar2025}. Its large collecting area, optimized antenna design, and precise calibration strategies will allow unprecedented access to the redshift range z $\sim$ $6-27$, corresponding to the formative billion years of cosmic history \citep{Acedo2017,2019Garc, 2015Jones}.

In parallel, the community is exploring hybrid approaches that combine classical statistical tools with modern computational frameworks. Techniques such as deep learning, convolutional neural networks (CNNs), and simulation-based inference have already demonstrated success in classifying fast radio bursts (FRBs), identifying radio frequency interference (RFI), and detecting H II regions during reionization capabilities that can be adapted for 21-cm foreground mitigation \citep{2020Agarwal, 2021Bianco, 2020Hassan}. In addition, end-to-end simulations are being developed to test and validate these techniques, integrating modules for signal modeling, foreground subtraction, calibration systematics, and power spectrum estimation across instruments such as MeerKAT and SKA-Low. 


Instrumental artifacts, such as cable reflections, bandpass irregularities, and spectral leakage, pose another major obstacle. Comprehensive analyses of these effects are underway to quantify and mitigate their impact on spectral smoothness and dynamic range, ensuring reliable signal recovery in future SKA1-Low observations \citep{Acedo2017,2019Garc, 2015Jones, 2019ApJ...875...70B}. Accurate calibration across the full frequency range remains a critical requirement for minimizing leakage of foreground power into the EoR window.

Beyond first detection, the SKAO’s ultimate goal is to produce the first high signal-to-noise measurements of 21-cm brightness temperature fluctuations, providing both statistical and tomographic maps of the IGM evolution during the first billion years \citep{2015Pritchard, 2020Greig, 2015Mesinger}. These measurements will place strong constraints on the astrophysical processes driving reionization, the thermal history of the IGM, and the underlying cosmological parameters. 

The chapter is organized as follows: In Section~\ref{sec:foreground_sources}, we describe different foreground sources and how systematic effects can exacerbate the foreground problem. In Section~\ref{sec_mitigation}, discuss strategies for foreground subtraction techniques, including avoidance and source subtraction. In Section~\ref{sdc3a}, we review the results from the SDC3a challenge, current challenges and limitations in foreground subtraction, and an outlook on future directions. We finalize our conclusion in Sec~\ref{conclusion}

\section{Foregrounds sources}\label{sec:foreground_sources}
This section will provide a detailed description of Galactic and Extragalactic Foregrounds, Recombination Lines (RRLs), including a survey of codes for simulating foregrounds and a discussion of how systematic effects can exacerbate the foreground problem.

At low frequencies, foreground sources are dominated by galactic and extragalactic synchrotron emission, which is 4-5 orders of magnitude brighter than the 21-cm signal \citep{1999A&A...345..380S,di200421, 2005ApJ...625..575S}. This makes detecting the faint cosmological 21-cm signal extremely challenging. Foreground sources include diffuse synchrotron emission \citep[DSE;][]{1999A&A...345..380S}, free-free emissions \citep{2004ApJ...606L...5C}, extragalactic point sources \citep{2002ApJ...564..576D}, etc.  Recent studies reveal that instrumental chromaticity \citep{Kern2020}, frequency-dependent beam patterns \citep{Chokshi2024}, and calibration errors \citep{2017MNRAS.470.1849E, 2019ApJ...875...70B} can imprint frequency-dependent structure onto otherwise spectrally smooth foregrounds and obscure foreground-free 21-cm dominated modes. Therefore, studying and developing all-sky foreground models through observations with SKA pathfinders (LOFAR, MWA, HERA, and uGMRT) is crucial for mitigating foreground contamination in future SKA observations.
\begin{figure}
    \centering
    \includegraphics[width=5in,height=3in]{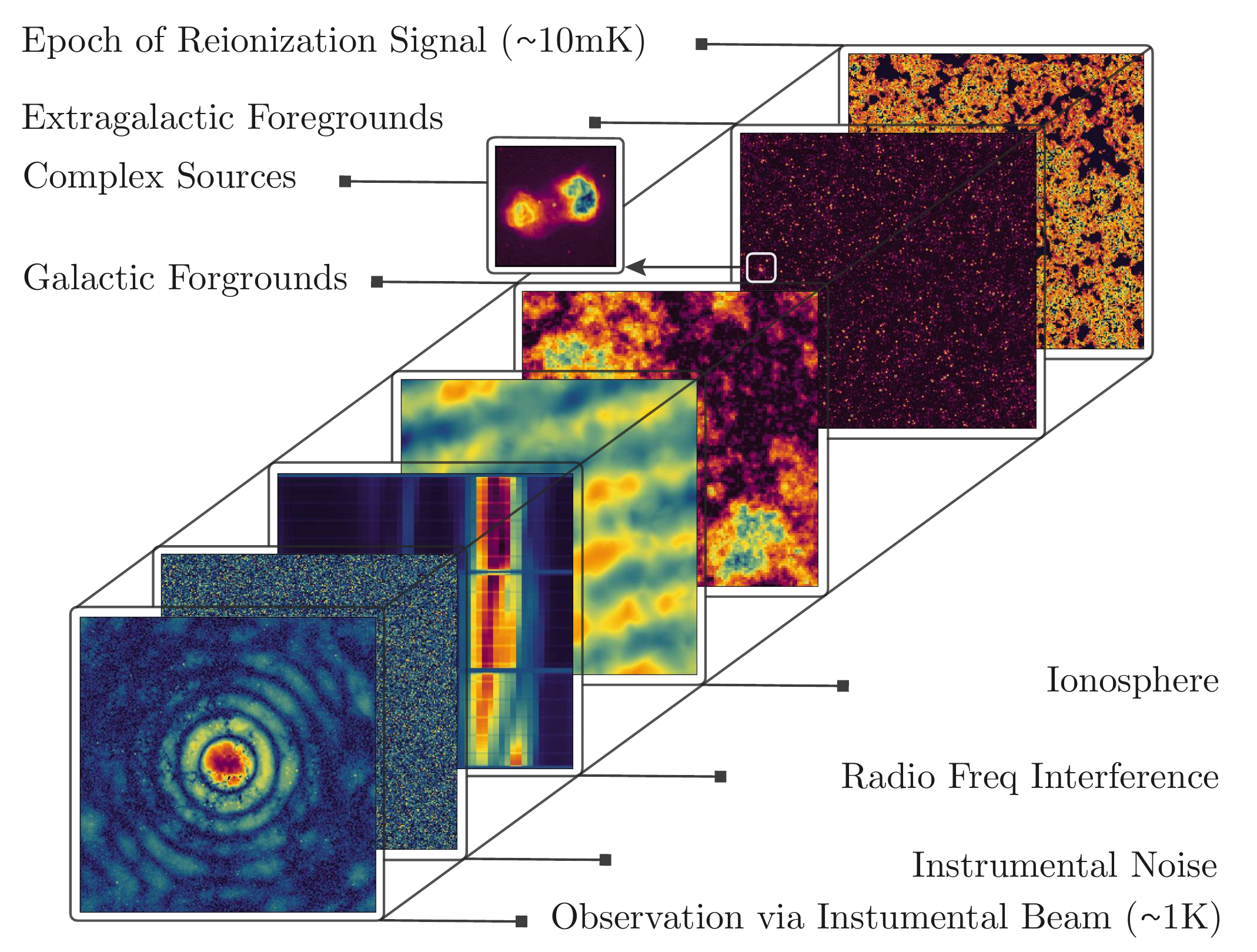}
    \caption{The pictorial representation of the key components of challenges in CD/EoR observations, from the faint 21-cm signal to foreground, terrestrial, and instrumental effects \citep{Chokshi2024}}
    \label{fig:Fig1}
\end{figure}
\subsection{Galactic foregrounds}\label{sec:galactic_foregrounds}
Galactic foregrounds include diffuse Galactic synchrotron emission (DGSE) in total intensity, polarized diffuse emission (Faraday-rotated), and galactic free-free emission \citep{2020MNRAS.496.1232L}, however, the free-free emission is sub-dominant at low-frequency. 

\subsubsection{Galactic emission in total intensity}\label{sec:gal_total}

The DGSE is mainly produced by the electrons spiralling in the Galactic magnetic field lines \citep{rybicki1979}. We require a precise characterization and a detailed understanding of the DGSE to remove foregrounds in 21-cm experiments reliably. Additionally, the angular fluctuations of the DGSE can be used to probe the fluctuations in the magnetic field and also the fluctuations in the cosmic ray electron density of the turbulent interstellar medium (ISM) of our Galaxy \citep{Cho2008,Lazarian2012,Iacobelli2013}. \citet{Lazarian2012} outlined the directions of how synchrotron foreground emission can be separated from the cosmological signal, i.e., from cosmic microwave background or highly redshifted HI 21-cm emission, which is relevant for EoR studies.

The angular power spectrum $(C_{\ell})$ of the DGSE quantifies the statistical fluctuations in the sky. The $C_{\ell}$ of the DGSE in total intensity fluctuations spanning over large portions of the sky can be modelled by a power law i.e. $C_{\ell}\propto\ell^{-\beta}$ \citep{Tegmark2000,Bacci2001}. \citet{Laporta2008} have analysed the 408-{\rm MHz} Haslam map \citep{Haslam1981,Haslam1982} and the 1420-{\rm MHz} survey data \citep{Reich1982,Reich1986,Reich2001} separately to measure the $C_{\ell}$ of the DGSE and found $\beta$ values in the range $2.6 - 3.0$ down to the angular multipoles  of $\ell = 200$ and 300 at 408 and 1420 {\rm MHz} respectively. These studies also show that $\beta$ steepens (or increases)  towards higher Galactic latitude.

The $C_{\ell}$ of the DGSE are not well quantified at the frequencies and angular scales relevant for detecting the cosmological 21-cm signal from the EoR. \citet{Parsons2010} have presented the all-sky synthesized map and estimated the $C_{\ell}$ between 139 {\rm MHz} and 174 {\rm MHz}. It has also been measured in only a few small fields at low Galactic latitude $|b| <
14^{\circ}$ in the frequency range 150 - 160 {\rm MHz} \citep{Bernardi2009,Bernardi2010,Ghosh2012,Iacobelli2013,Choudhuri2017}. \citet{Bernardi2009} and \citet{Ghosh2012} have, respectively, analysed 150 {\rm MHz} WSRT and GMRT observations and extracted the values of $\beta=2.2 \pm 0.3$ and $\beta=2.34 \pm 0.28$ up to $\ell=900$. \citet{Iacobelli2013} have measured the $C_{\ell}$ of the DGSE at 160 {\rm MHz} using LOFAR data and reported that the angular power spectrum has a slope $\beta \approx 1.8$ down to the angular multipoles $\ell$ of 1300. \citet{Choudhuri2017} have analysed two fields from the TIFR GMRT Sky Survey (TGSS)-ADR1 survey at 150 {\rm MHz} \citep{2017A&A...598A..78I} and measured the $C_{\ell}$ of the DGSE across the $\ell$ range $240 \le \ell \le500$ and found that the values of $\beta$ are $2.8 \pm 0.3$ and $2.2 \pm 0.4$ respectively in the two fields. Earlier, \citet{Chakrabprty2019} have measured the $C_{\ell}$ in ELAIS-N1 field and found the $\beta$ values consistent with earlier measurements. Recently, \citet{2025arXiv251102375S} have measured the $C_{\ell}$ in ELAIS-N1 field in wide-band  from $120-500$\,MHz and found the spectral break at  $\nu = 230\,{\pm}\,5$\,MHz with synchrotron age $t_\text{syn} = 106\,{\pm}\,1$\,Myr of cosmic ray electrons (CRe). This indicates a low energy cut off in CRe population at low frequencies. All of these results are restricted to a small portion of the sky $\le 6^{\circ} \times 6^{\circ}$.

\citet{2020MNRAS.494.1936C} used the TGSS data \citep{Sirothia2014,2017A&A...598A..78I} to quantify the statistical properties of the DGSE, which is expected to be the dominant foreground contribution after source subtraction. The TGSS contains observations in $5336$ pointings covering a large fraction ($90 \%$) of the total sky in the declination range $\delta >-55^{\circ}$. They found that amplitude of the measured $C_{\ell}$ falls significantly after subtracting the point sources, and it is also slightly higher in the Galactic plane for the residual data. They conclude residual $C_{\ell}$ is most likely to be dominated by the Galactic synchrotron emission, and the amplitude of the residual $C_{\ell}$ falls significantly away from the Galactic plane. The DGSE measurements in the total intensity are quite symmetric in the Northern and Southern hemispheres except in the latitude range $15-30^{\circ}$ which is the transition region from the disk dominated to diffuse halo dominated region.

\subsubsection{Polarized Galactic emission}\label{sec:polgal}
Galactic synchrotron emission is not only the dominant foreground contaminant in total intensity at the frequencies targeted by 21 cm experiments, but it is also intrinsically linearly polarized. As radiation propagates through the magnetized and ionized interstellar medium, its polarization angle is rotated by the Faraday rotation \citep[e.g.][]{burn1966}, an effect that scales with the square of the wavelength. The cosmological 21 cm signal is expected to be unpolarized; however, instrumental imperfections introduce polarization leakage, which couples polarized foregrounds into total intensity. This leakage is particularly problematic because the Faraday-rotated polarized signal exhibits a strong frequency-dependent structure, which can mimic or obscure the cosmological signal \citep{jelic10, moore13, nunhokee17, spinelli19}. Accurate calibration and detailed characterization of instrumental polarization \citep{sutinjo15, asad15, asad16, asad18}, together with deep polarization sky surveys, are therefore essential prerequisites for 21 cm cosmology.

Early attempts to probe diffuse polarized Galactic emission around 150 MHz were made with the Giant Metrewave Radio Telescope \citep{pen09} and the WSRT \citep{bernardi09, bernardi10}. However, the full complexity of the polarized sky only became evident with the advent of LOFAR and MWA, which revealed striking morphologies of polarized structures \citep{bernardi13, jelic14, jelic15, lenc16} through the application of rotation measurement (RM) synthesis \citep{brentjens05}. More recently, the LOFAR Two-meter Sky Survey (LoTSS) has delivered the most extensive low-frequency polarization survey to date, covering a large fraction of the northern sky \citep{erceg22, erceg24}. These results indicate that Faraday structures with potential impact on the cosmological 21 cm signal — namely emission detected at Faraday depths $|\Phi|\geq 15~{\rm rad~m^{-2}}$ \citep{chapman19} — are relatively rare and generally of low brightness (Ceraj et al., in prep.), suggesting that moderate levels of polarization leakage are unlikely to pose a serious source of contamination. Consistent with this, \citet{snidaric23} presented the first deep polarimetric study of Galactic synchrotron emission at low radio frequencies by stacking 21 LOFAR observations of the ELAIS-N1 field and detected only a very faint additional diffuse polarized component at large Faraday depths relevant for cosmological experiments. The weakness of this component is explained by strong differential Faraday rotation and depolarization effects along the line of sight, which significantly reduce the observable polarized intensity at such Faraday depths.

\subsection{Extragalactic foregrounds}\label{sec:extra-galactic}
Extragalactic foregrounds include all radio sources beyond the Milky Way that contaminate 21-cm observations \citep{Shaver1999,DiMatteo2002,Jelic2008}. Key contributors are radio-loud AGN and star-forming galaxies, with smaller contributions from diffuse cluster emissions \citep{Wilman2008,Bonaldi2019}. These foregrounds dominate small angular scales and account for $\sim 30\%$ of total foreground power at 150 MHz \citep{Jelic2008,Bowman2009}. They are 3–4 orders of magnitude brighter than the cosmic 21-cm signal in total intensity, with fluctuations exceeding the EoR signal by 2–3 orders of magnitude \citep{Shaver1999,DiMatteo2002,Bowman2009}. This section summarizes extragalactic foreground components, their statistical characterization, and implications for SKA-Low.


\subsubsection{Components and Phenomenology}\label{sec:componets_phn}
Extragalactic Point Sources: Radio-loud AGN (radio galaxies and quasars) are the dominant extragalactic contaminants  \citep{Wilman2008,Bonaldi2019}. They often exhibit steep power-law spectra ($S_\nu \propto \nu^{\alpha}$ with $\alpha\sim -0.7$) and high luminosities, sometimes reaching hundreds of Jy at low frequencies  \citep{HurleyWalker2017,Murray2017,Bonaldi2019}. About $\sim 10\%$ have flatter spectra or spectral turnovers (e.g. gigahertz-peaked sources) that depart from smooth power laws, with some exhibiting turnovers at low frequencies ($\lesssim100$–200 MHz; e.g. \citealt{Callingham17, 2019MNRAS.485.2447M}) and even approaching the canonical synchrotron self-absorption limit \citep{2019MNRAS.485.2447M}. Many AGN host extended lobes that are mostly unpolarized, but their compact cores or jets can exhibit a few percent polarization. However, strong Faraday depolarization at $<200$ MHz means only a very small fraction of sources show detectable polarized emission – for example, deep LOFAR 150 MHz observations of 16 deg$^2$ detected only ten polarized sources (surface density $\sim0.6$ per deg$^2$) \citep{Asad2016,Spinelli2018}. 

Star-Forming Galaxies: Normal galaxies with vigorous star formation host diffuse synchrotron emission from cosmic-ray electrons in their disks and are much fainter (typically mJy at 150 MHz) but far more numerous than bright AGN \citep{Shimwell2019,Best2023}.
Their spectra are also steep ($\alpha\sim-0.7$ to $-0.8$) and largely unpolarized \citep{HurleyWalker2017,Bonaldi2019}.  Below $\sim1$ mJy, they dominate source counts, producing cumulative confusion noise that manifests as a diffuse extragalactic background \citep{Shaver1999,Jelic2008}.


Diffuse Cluster Emission: Galaxy clusters can host megaparsec-scale radio halos and relics powered by intracluster shocks and turbulence \citep{Jelic2008,Spinelli2018}. These are extended (arcminute-scale) sources with very steep spectra ($\alpha\lesssim -1$). They are relatively rare and contribute negligibly to the overall point-source counts, but a bright halo or relic in an EoR field would appear as a patch of diffuse extragalactic foreground \citep{Jelic2008,Li2019}. Such emissions are unpolarized or weakly polarized. 


Phenomenology: Across these components, extragalactic foreground spectra are predominantly smooth \citep{DiMatteo2002,Bowman2009,Morales2012}, a fact exploited in many foreground mitigation schemes \citep{Harker2010,Chapman2012,Chapman2013}. Polarization is generally low, but  frequency-dependent Faraday rotation of polarized emission can imprint small-scale spectral structure in Stokes $I$ measurement \citep{Asad2016,Spinelli2018}.   Bright radio AGN show mild angular clustering on degree scales, while star-forming galaxies have weaker clustering \citep{Murray2017,Shimwell2019}.
Finally, the extragalactic foreground sky is dynamic range–challenging: a typical deep EoR field may contain a few $\sim10$ Jy AGN, hundreds of $\sim0.1$–1 Jy sources, and tens of thousands of fainter sources, all of which must be handled in calibration and imaging \citep{Procopio2017,Murray2017}.

\subsubsection{Statistical Characterization and Priors}\label{sec:stats_charac}
Precision modeling of extragalactic foregrounds relies on statistical priors from radio surveys. The source flux distribution $dN/dS$ at low frequencies has been quantified by MWA-GLEAM and LOFAR surveys. GLEAM (72–231 MHz) provided all-sky counts down to $S_{154\text{MHz}}\approx50$–$100$ mJy, with sources above 0.5 Jy following a near power-law and average spectral index $\alpha\approx -0.8$ \citep{HurleyWalker2017}. LOFAR deep fields extended counts to sub-mJy levels, revealing a rise from emerging star-forming galaxies and complex population transitions around $\sim1$–2 mJy that current models struggle to reproduce \citep{Shimwell2019,Best2023}. Recent recalibrations using LoTSS data predict fewer faint sources and lower small-scale point-source power than earlier prescriptions like T-RECS \citep{Lin2024}. 

The spectral index distribution of extragalactic sources is also an important prior.The spectral index distribution is typically $\alpha \approx -0.7$ to $-0.8$, with a minority of flatter-spectrum compact cores and ultra-steep sources \citep{HurleyWalker2017,Bonaldi2019}. Models often assume Gaussian or double-population distributions of $\alpha$, calibrated by multi-frequency surveys. 

Another statistical input is the two-point angular clustering of sources.  In continuum surveys, bright radio AGN show mild clustering on degree scales (correlation amplitude $\sim10^{-3}$–$10^{-2}$), reflecting their association with massive halos. Star-forming galaxies have weaker clustering. In foreground simulations these effects are sometimes included (e.g. by Poisson sampling sources onto large-scale structure frameworks, or via measured angular correlation functions) \citep{Wilman2008,Murray2017}.

Finally, polarization statistics serve as priors for leakage modeling. Polarization statistics indicate only $\sim1$–2\% of bright sources have detectable polarized flux above $\sim0.1$\% at 150 MHz, justifying treating Stokes $I$ foregrounds as unpolarized to first order \citep{Asad2016,Spinelli2018}.

\subsubsection{Implications for SKA-Low}\label{sec:implification_forska}
Modelling and Calibration Requirements: Achieving the EoR science goals with SKA-Low demands extremely accurate foreground modelling and subtraction \citep{Chapman2019,Bonaldi2019}. The SKA-Low will have the sensitivity to detect $\sim10^7$ discrete sources across its band, essentially every foreground source down to $\mu$Jy levels \citep{Bonaldi2019,Murray2017}, which will be used as a sky model for initial calibration \citep{Lin2024}. Simulations indicate that in-field sources need to be modelled down to at least $\sim 1-10$ mJy to prevent significant unmodeled confusion noise in deep integrations \citep{Procopio2017,Murray2017}. 
Bright out-of-field sources ( > 1 Jy), can be picked up by the wide station beam and should be  “de-mixed” or subtracted via specialized calibration solutions \citep{Procopio2017}.

The dynamic range required is formidable, and the calibration and imaging must suppress foregrounds (up to tens of thousands of K in brightness) to the $\sim$mK level of the 21-cm signal, implying  calibration errors must be kept below $\sim10^{-5}$ (0.001\%) in relative terms \citep{datta2010bright, Vedantham2012,Morales2012}. Solutions include using bright source “peeling” to subtract the brightest few hundred sources individually, applying direction-dependent beam corrections, and iteratively refining sky models as deeper sources are revealed \citep{Procopio2017,Asad2016}.

In summary, SKA-Low will push foreground mitigation into a regime of precision subtraction and careful error control. Requirements include deep source catalogs (possibly leveraging SKA-Low itself to create them), high dynamic range imaging (innovations in deconvolution to handle confusion), and extremely well-characterized instrumental responses (to inform physics-based corrections for beam chromaticity and polarization leakage) \citep{Procopio2017,Asad2016}. By meeting these requirements, SKA-Low can approach the theoretical thermal-noise limit after foreground cleaning, opening the path to a high-fidelity detection of the 21-cm signal.

\subsection{Ionosphere}\label{sec:ionosphere}
The Earth's ionosphere is a known source that creates an obstacle to low-frequency radio astronomical imaging. It is a partially ionized layer of turbulent plasma, composed of turbulent disturbances of various spatiotemporal length scales,  extending from about 50 to 1000 km in altitude, created by solar radiation. This inhomogeneous plasma density fluctuations cause refraction, diffraction, scintillation, and absorption of incoming wavefronts, introducing direction- and time-dependent effects that limit the dynamic range of interferometric imaging. These effects become more pronounced for observations involving longer baselines of the interferometer and are particularly challenging to calibrate in the low signal-to-noise ratio (S/N) regime. Therefore, direction-dependent calibration must be performed at shorter time intervals that match the ionospheric variability. The dynamic nature of the ionosphere produces flicker noise ($1/f$ noise, where $f$ is the dynamical frequency) \citep{Datta2016}. Since these errors do not reduce with long integration. This rapid variation of the uncorrectable ionospheric phase fluctuations introduces speckle noise, i.e., halo structure around point sources \citep{Koopmans2010}.

There are several direction-dependent calibration techniques (see Section~\ref{dd-calibration} for more details and another application) have been developed to mitigate ionospheric errors \citep{2009A&A...501.1185I, 2018A&A...615A.179D,  2018MNRAS.478.2337R, Albert2020, 2022MNRAS.510.2718Y,  2021A&A...648A...1T, 2023ascl.soft05008T}. Generally, calibration is performed by estimating the Jones matrices under the assumption of a perfectly known sky model. The accuracy of this technique, however, is limited by the signal-to-noise ratio (SNR) and the number of directions or sources considered during the calibration process. The limitations of traditional calibration result in residual direction-dependent errors (DDEs) or ionospheric offsets that primarily constrain the upper limits of the CD/EoR power spectrum measurements \citep{Pal2025}. Therefore, mitigating these residual DDEs is essential for a robust interpretation of the faint CD-EoR signal in both the Fourier and image domains. There is a need for the development of efficient ionospheric calibration algorithms for the upcoming SKA-Low observations to extract the 21-cm power spectra from the CD/EoR. For machine learning applications to ionospheric error mitigation, we refer the interested reader to the \citet{Acharya02.2026.SKA}.

\subsection{Radio Frequency Interference}\label{sec:rfi} 
Some of the strongest sources of radio waves are not from astrophysical sources, but radio and TV broadcasts, high-speed wireless communications (e.g. cell phone networks and WiFi), and radar. Radio interferometers are often built in remote and radio-quiet locations to avoid these sources of Radio Frequency Interference (RFI). However, transient radio sources in the sky, such as satellites, are much more difficult to avoid. The number of active satellites has rapidly increased (from 1,000 in 2013 to over 5,000 in 2022~\citep{GAO-22-105166}), becoming a growing concern in the radio astronomy community.\\
Satellite operators are required to use only the frequency bands that they have registered with the International Telecommunication Union (ITU), and certain frequency bands are dedicated to radio astronomy to avoid interference with essential observations. Yet despite these protections, satellites may emit RFI on frequencies outside their assigned bands. Interference caused by unintended emissions from Starlink satellites has recently been detected by the LOFAR telescope in the range of 110-188 MHz \citep{lofarRFI} and below 100 MHz with NenuFAR~\citep{Zhang_2025}, far outside Starlink's assigned transmission frequency of 10.7–12.7 GHz. This Unintented ElectroMagnetic Radiation (UEMR) can be generated from various components on satellites, such as communication systems and power supplies, and are not currently regulated by international standards. Recent studies, including those by LOFAR and the Atacama Large Millimeter/submillimeter Array (ALMA), have shown that astronomical research in general is becoming more difficult as a result of these emissions \citep{siringo2024spectrum,engelbrecht_radio_2024, wang_satellite_2021, 10907102}.

Even extremely faint RFI from satellite UEMR can overwhelm measurements of cosmological 21 cm~\citep{Wilensky_2023}, and must be mitigated, flagged, or subtracted.\\
The uGMRT Bands 2 (120–250 MHz) experiences both broadband impulsive RFI, primarily from power-line discharges, industrial equipment, and electric motors and narrowband persistent RFI arising from FM radio, television, and communication transmitters in the vicinity of the observatory \citep{2023JApA...44...37B}. These signals often exceed the system thermal noise by 30–40 dB, thereby contaminating visibilities and introducing non-Gaussian outliers in the temporal and spectral domains \citep{2016JAI.....541018B}. The spatially correlated nature of far-field RFI makes it particularly detrimental for interferometric arrays, as coherent contamination across antennas can masquerade as astronomical structure. The problem intensifies at low frequencies due to increased sky temperature, ionospheric propagation of terrestrial signals, and wider observing bandwidths that enhance susceptibility to contamination \citep{2025arXiv250617131K}. Continuous RFI monitoring at the uGMRT site has revealed temporal and seasonal variation in interference, often linked to monsoon humidity and wind-induced arcing on power lines \citep{2010MNRAS.405..155O, 2015PASA...32....8O, 2022JAI....1150008B}. Understanding this dynamic RFI landscape is therefore critical, not only for real-time data protection but also for designing calibration and foreground-avoidance strategies compatible with future SKA-Low observations \citep{2004ExA....17..261E, 2023JApA...44...28G}.

\subsection{Foregrounds simulations code}\label{sec:fcode}
Simulation code is used to generate synthetic observations that mimic real data, allowing testing of calibration algorithms, imaging pipelines, and science analysis in a controlled setting. The scope of simulations ranges from high-fidelity reproductions of the sky and instrument (for end-to-end pipeline verification) to simplified scenarios for analytical studies.
Simulations help quantify performance (e.g. sensitivity, image fidelity) under various conditions, optimize survey strategies, and identify potential systematic errors in advance. 
In essence, simulation code provides a virtual SKA, ensuring that both hardware design and data processing software can meet the SKA’s requirements.

\subsubsection{Sky Models for Simulations}\label{sec:sim_sky}
Realistic sky modelling is a critical component of simulations, combining astrophysical signals of interest with foreground emission across relevant frequency bands. For diffuse Galactic emission, empirical all-sky models such as the Global Sky Model (GSM) \citep{deOliveiraCosta2008} and Global MOdel for the radio Sky Spectrum (GMOSS) \citep{SathyaRao2017} are widely used. The GSM provides frequency-dependent sky brightness by combining radio surveys and has extended up to $\sim$5~THz with principal component analysis. The GMOSS is a physically-motivated alternative that fits low-frequency spectra using components for synchrotron, free-free, and thermal emission.

In addition to diffuse models, simulators ingest discrete source catalogues or sky catalogues from surveys such as NVSS and GLEAM (see Section~\ref{sec:extra-galactic} for more details of extragalactic radio sources). \texttt{fg21sim} \citep{li2019separating, Shan2024} is widely used for modelling low-frequency foregrounds, generating full-sky spectral cubes that integrate seamlessly with simulators such as OSKAR or pyuvsim. In summary, sky models in simulations are assembled from a combination of analytic or empirical diffuse components, spectral line signal models, and point source catalogues, covering the wide range of scales and frequency behaviour expected in real data.

\subsubsection{Instrument and Propagation Effects}\label{sec:error_model}
To produce realistic observables, simulation code must incorporate detailed instrument and propagation effects. The fundamental basis for most radio interferometer simulations is the Radio Interferometer Measurement Equation (RIME) \citep{Hamaker1996}, a matrix formulation that describes how the sky brightness is transformed into complex visibilities by the instrument's response. We refer the reader to \citet{Smirnov2011} for a more detailed explanation of the RIME formalism. Direction-independent effects such as complex gain drifts or bandpass shapes are typically represented by per-antenna Jones terms, whereas direction-dependent effects (DDEs) like beam shape or ionospheric phase screen require more complex modelling across the field of view \citet{Mazumder2022, Pal2025, 2025Tripathi}.

Wide-field imaging presents challenges due to the non-coplanar nature of arrays such as SKA, LOFAR, and NenuFAR, which introduces a $w$-term in the interferometric phase, distorting the observed visibilities. Simulation codes account for this either by direct spherical geometry calculations of visibilities or by using techniques like $w$-projection in the imaging stage. Thermal noise can be injected based on the system temperature ($T_{\rm sys}$) and integration time, often modelled as Gaussian random visibilities with a variance set by the system-equivalent flux density (SEFD) of the telescope. The ionosphere is another critical factor for low-frequency SKA simulations: time-variable, spatially correlated phase delays and Faraday rotation can be applied via phase screen models or more advanced ray-tracing methods to assess their impact on calibration and imaging. In summary, a high-fidelity simulation utilises the RIME as a backbone to apply a cascade of effects, from the sky to voltage beam, to antenna gains, and to the correlator, producing mock visibilities that closely approximate what the real SKA will observe.

\subsubsection{Core Simulation Tools}\label{sec:tools}
Over the last decade, several simulation tools have been developed for SKA studies. Each has different strengths and focuses, and is widely adopted by the community.
Table~\ref{tab:sim-compare} summarizes several key characteristics. OSKAR\citep{Mort2010} is a well-established simulator designed originally for SKA-Low. Written in C++ with GPU/CUDA acceleration, it can efficiently simulate visibilities for large antenna arrays and is built to handle beamforming for aperture arrays. It supports full polarization and implements the RIME (including direction-dependent gains for station beams) to generate realistic SKA-Low observations. There are other simulation tools used to forward model the sky signal through the instrument,, including WODEN \citep{Line2022}, PRISim \citep{PRISimASCL}, and Pyuvsim \citep{Lanman2019}. We refer the reader to the \citet{deLeraAcedo01.2026.SKA} for a comprehensive overview of end-to-end simulations and pipeline validation.

\begin{table}[t]
\centering
\small
\setlength{\tabcolsep}{6pt}
\caption{Major simulators at a glance (compact view). Columns use fixed widths to improve fit in single-column layouts.}
\label{tab:sim-compare}

\begin{tabular}{p{0.30\linewidth} p{0.20\linewidth} p{0.40\linewidth} p{0.10\linewidth}}
\hline\hline
\textbf{Simulator} & \textbf{Impl./Acceleration} & \textbf{Sky support (typical)} & \textbf{Pol.}\\
\hline
OSKAR \citep{Mort2010,OSKARDocs} & C++ with CUDA; Python bindings &
Point + diffuse via station/element beams; large-$N$ arrays & Full\\

RASCIL \citep{RASCILDocs,SensCalcRASCIL} & Python/NumPy; Dask workflows &
Point, images; reference sim/cal/imaging pipelines & Full\\

WODEN \citep{Line2022} & C++/CUDA (GPU) &
Discrete all-sky; fast low-freq workloads & Full\\

pyuvsim \citep{Lanman2019} & Python + MPI (CPU clusters) &
Point + diffuse (HEALPix) with exact RIME eval. & Full\\

PRISim \citep{PRISimASCL} & Python (vectorized/MPI) &
Point, gridded diffuse; EoR pipelines & $I$\\

MeqTrees \citep{Noordam2010} & C++ back-end + Python front-end &
Arbitrary RIME trees; DDE experiments & Full\\
\hline
\end{tabular}
\end{table}

\section{Current mitigation Strategies}\label{sec_mitigation}
This section will discuss strategies for foreground avoidance, source subtraction techniques (including direction-dependent calibration and sky-model subtraction), and residual foregrounds subtraction methods such as Gaussian Process Regression (GPR), Generalized Morphological Component Analysis (GMCA), Principal Component Analysis (PCA), and their variant algorithms, as well as recent machine learning approaches. We will include a review of how systematic effects can be mitigated, drawn upon the experience from SKA precursors.

\subsection{Avoidance methods: the foregrounds wedge}\label{sec:compare_tech}
The avoidance method leverages on the characteristic differences between astrophysical foregrounds and the cosmological 21 cm signal in Fourier space. Foregrounds, which are typically smooth in frequency, tend to dominate at low line-of-sight wavenumbers. Additionally, interferometers are chromatic instruments, resulting in a mixture of spatial and spectral modes in the cylindrical power spectrum, a phenomenon known as mode mixing \citep{datta2010bright,liu2011method}. Longer baselines are more chromatic, resulting in a characteristic wedge-like region in the cylindrical power spectrum within which flat-spectrum foregrounds are expected to be confined. The maximum extent of the foreground wedge is defined by the horizon delay line \citep{Vedantham2012, Morales2012,trott2012impact,thyagarajan2013study}. In the delay power spectrum approach, the wedge is a direct consequence of using baseline lengths as proxies for the spatial scale \citep{parsons2012per}, and the horizon delay line represents the light-crossing time corresponding to a given baseline length. In the reconstructed power spectrum approach, mode-mixing arises due to imperfect $uv$ coverage \citep{murray2018effect}, and the horizon line is dependent on the elevation of the phase center \citep{munshi2025beyond}.

In contrast, the EoR signal, which contains more spectral structure, extends into higher line-of-sight modes. This separation creates a relatively clean region, referred to as the EoR window \citep{liu2014epoch1,liu2014epoch2}, where the contribution from foregrounds is significantly reduced, allowing for a more reliable extraction of the underlying cosmological signal. In practice, any spectral structure imprinted on the foregrounds by instrumental effects results in leakage of foreground power into the EoR window. The extent of foreground leakage also depends on the window function used in the Fourier transformation along frequency. Foreground avoidance approaches thus typically also includes a buffer region beyond the horizon line to avoid the modes in the cylindrical power spectrum strongly affected by spectral leakage.

\subsection{Visibility-correlation approach}
\label{sec:viscorr}
An alternative approach to estimate 21-cm statistics is to estimate the `relevant' quantities in the `frequency' space first, before going to the Fourier space. This approach is traditionally referred to as a visibility-correlation approach \citep{2001JApA...22...21B, 2005MNRAS.356.1519B}. A two-visibility correlation directly yields the key quantity of interest -- the multi-frequency angular power spectrum (MAPS) $C_\ell(\nu_1, \nu_2)$, which contains the entire two-point statistics of the signal. If the signal is assumed to be ergodic, then $C_\ell(\nu_1, \nu_2) = C_\ell(\nu_1 - \nu_2)$, is the separation between two frequencies, and the cylindrical power spectrum $P(k_\perp, k_\parallel)$ is the Fourier transform of $C_\ell(\Delta\nu)$ along $\Delta\nu$. In MAPS $C_\ell(\Delta\nu)$, the foreground wedge appears as an $\ell$-dependent oscillatory feature along $\Delta\nu$ -- the larger the angular scale, the faster the oscillation \citep{2011MNRAS.411.2426G} -- which in turn gives rise to the wedge in $P(k_\perp, k_\parallel)$ \citep{2022MNRAS.516.2851P}. \cite{2022MNRAS.516.2851P} further illustrates this using a simplified model with a single point source foreground for which $C_\ell (\Delta\nu) \propto \cos{(\ell \theta \Delta \nu/\nu_c)}$ where $\theta$ is the sine of the angle between the source position and the phase center of the observation. The extent of the foreground wedge is determined by the position ($\theta$) of the wide-field source,  which can maximally reach the horizon limit $\theta \sim 1$. Note that the oscillations fundamentally originate due to baseline migration; however, their amplitude can be mitigated by suppressing the sidelobe response of the telescope \citep{2011MNRAS.418.2584G}. This idea gave rise to the Tapered Gridded Estimator\,(TGE; \citealt{samir14, samir16}). We will discuss how foregrounds and several systematics can be uniquely treated in the following sections using this visibility-correlation approach of TGE.

A direct extension to three-point statistics, three-visibility correlations estimate the multi-frequency angular bispectrum and its Fourier bispectrum counterpart \citep[for more details, see \citet{deLeraAcedo01.2026.SKA}]{2005MNRAS.358..968B, 2025arXiv250704964S}. The bispectrum is estimated using closed triangle configurations formed by three modes. The estimated bispectrum also exhibits the foreground wedge features that are very similar to the power spectrum, but the structure of the wedge is far more complex due to the involvement of three $k$-modes. We can also identify a clear EoR window in the bispectrum space, where all three triangle sides are outside the foreground wedge. For a particular case with MWA, around $83.6\%$ of triangles are found to lie within the EoR window. Although this is highly promising to probe the non-Gaussian nature of the EoR signal, substantial leakage into the EoR window due to the periodic pattern of missing frequency channels in the MWA bandpass makes it challenging to obtain reliable constraints. The prospect is expected to be better for SKA-Low which will have a better bandpass and thereby yield many more clean triangle configurations. At present however, further progress in this direction demands the removal of the dominant foreground components through the methods described below.

\subsection{Sky source subtraction}\label{sec:source_subtraction}
Foreground subtraction is a well-established technique in which the foreground emission is modelled and subsequently removed from the data. In this approach, bright sources are modelled and subtracted directly from the visibility domain, the space in which interferometric measurements are made, and this method has proven highly effective \citep{Morales2012, Pober2016}. However, in widefield observations, variations in the primary beam shape and ionospheric conditions introduce spatially dependent errors in the data. To mitigate these effects, several methods have been developed that simultaneously subtract sources along with their corresponding calibration solutions or corrections \citep{2008ISTSP...2..707M, Bernardi2011, Jordan2017, Procopio2017, Jordan2025}. These corrections has shown improvements in the resulting power spectra, with lower contamination into the wedge region \citep{Nunhokee2025, Mertens2025}.

The following subsections discuss the sky-model building approaches adopted by the 21 cm community, as well as the application of direction-dependent corrections for bright sources in the field to address ionospheric effects \citep{Jordan2017, Brackenhoff2024}.

\subsubsection{Sky-model building}\label{sec:sky_model_building}
For foreground subtraction, it is crucial to have a good foreground model to subtract the visibilities from given the faint nature of the 21 cm signal. Residuals from the subtraction process are prone to introduction spurious spectral structures to the power spectra \citep{Morales2012, Procopio2017, Barry2018}. Further, \cite{Pober2016} further demonstrated that incorporating sources beyond the main lobe of the primary beam into the sky model can significantly reduce contamination within the EoR window. Nevertheless, the brightest sources are often resolved at angular scales below 2~arcminutes, and inadequate modelling of these sources can lead to substantial subtraction errors \citep{Procopio2017}, thereby overwhelming the faint 21~cm signal. 

The MWA generally adopts a foreground subtraction approach, in which the sky model is constructed from several components of the southern sky, including the LoBES catalogue \citep{Lynch2021} and the GLEAM survey \citep{Wayth2015, HurleyWalker2017}. The sources are modeled using either a spectral-law or a curved power-law energy distribution. In addition, extended sources such as Fornax A, Centaurus A, and Galactic supernova remnants (SNRs) are incorporated into the model. Modeling these extended sources is particularly challenging due to the high level of accuracy required to achieve the desired noise level.

\citet{Line2020} implemented a shapelet-based model of Fornax A, which improved the subtraction process and consequently enhanced the resulting power spectrum. Similarly, \citet{Cook2022} employed multicomponent 2D Gaussian models for Galactic SNRs and Centaurus A. In the northern sky, Cygnus A is the dominant extended source. \citet{Ceccotti2024} used a forced-spectrum technique to overcome the limitations in modelling Cygnus A, resulting in a significant improvement in the cylindrical power spectrum.

\subsubsection{Direction-dependent calibration}\label{dd-calibration}
Direction-dependent (DD) calibration became essential with wide fields of view and high dynamic range observations, where traditional direction-independent (DI) self-calibration fails to correct ionospheric distortions and complex primary beam shapes \citep{2009ASPC..407..375B, 2022MNRAS.510.2718Y}. The Source Peeling and Atmospheric Modeling (SPAM) technique for the uGMRT corrects ionospheric delays using bright calibrators distributed across the field \citep{2009A&A...501.1185I, 2014ASInC..13..469I, 2017A&A...598A..78I, 2025arXiv251102375S}, while LOFAR’s Facet Calibration partitions the sky into independent facets and calibrates each separately, enabling scalability \citep{2016ApJS..223....2V, 2018A&A...611A..87T, 2023ascl.soft05008T}. The DD calibration introduces a high number of free parameters, which can risk overfitting and suppress the faint 21-cm signal. To mitigate this, LOFAR  limits calibration to baselines longer than a cutoff, restricting any signal loss to that subset \citep{Mertens2025, Brackenhoff2024}. For further details on DD calibration techniques across different radio interferometers, we refer the reader to the \citet{deLeraAcedo01.2026.SKA}.

\subsection{Signal separation techniques}\label{sec:sigsep}
Even after bright sources have been subtracted and the wedge avoided, residual foregrounds and instrumental contaminants remain several orders of magnitude stronger than the cosmological signal. To recover the information in the foreground wedge, one must exploit differences in their statistical and spectral behaviour. Signal separation methods can be broadly grouped into three categories: (i) \emph{blind methods}, which decompose the data into statistically independent or sparse components without explicit astrophysical priors; (ii) \emph{Bayesian approaches}, which model the data with probabilistic priors on foreground smoothness, instrumental effects and the 21-cm covariance; and (iii) \emph{machine learning techniques}, which learn non-linear mappings directly from simulations or data. In practice, these approaches are often combined into hybrid pipelines, reflecting the need for robustness against both foreground complexity and instrumental systematics.

\subsubsection{Blind Source Separation methods}\label{sec:blind_source_sub}

Blind Source Separation (BSS) methods aim to disentangle the observed data into statistically independent or morphologically distinct components, without requiring detailed astrophysical priors. The motivation is that the foregrounds, although much brighter than the cosmological 21-cm signal, should occupy a smaller number of dominant modes in frequency or wavelet space, while the 21-cm signal and instrumental noise are less coherent. By exploiting this difference, BSS methods attempt to recover the cosmological signal in the residuals.

A first class of BSS techniques are based on statistical independence. Principal Component Analysis (PCA) removes the few largest eigenmodes of the data covariance, which are assumed to be dominated by foregrounds. While simple, this method risks partial signal loss if the cosmological modes overlap with foreground-dominated ones. Independent Component Analysis (ICA) extends this approach by maximizing non-Gaussianity of the components \citep{Chapman2012}. ICA assumes that the data can be expressed as a linear mixture of independent components, which are then separated by estimating the mixing matrix. In practice, ICA can capture more subtle spectral structures than PCA, but still suffers from leakage of 21-cm power into the reconstructed foregrounds.

A second class are methods exploiting morphological diversity. The most widely used example is the Generalized Morphological Component Analysis (GMCA) \citep{Bobin08, Chapman2013}. GMCA assumes that astrophysical foregrounds have sparse representations in a wavelet basis, while the 21-cm signal and noise remain incoherent in this space. This allows the algorithm to isolate foregrounds through their sparse coefficients. GMCA has been applied successfully to LOFAR-EoR observations, and was used in deriving an early LOFAR upper limits \citep{Patil2017}. However, later comparisons showed that GMCA struggles with complex instrumental mode-mixing, and does not reach the same performance as GPR in LOFAR data analysis \citep{Mertens2018}.

The strength of BSS methods lies in their flexibility and relative independence from astrophysical priors. They provide a powerful check against more model-dependent approaches. Their main limitations are the empirical choice of the number of components, sensitivity to noise and systematics, and the difficulty of quantifying uncertainties on the recovered cosmological signal \citep{Chapman2019}.

\subsubsection{Bayesian Approaches}\label{sec:bays_approach}
The Bayesian framework \citep{Ghosh:2015fxa, 2016ApJS..222....3Z, 2019MNRAS.488.2904S, 2023MNRAS.520.4443B} provides a principled approach for the joint inference of the cosmological 21-cm signal and astrophysical foregrounds, explicitly incorporating prior knowledge about their statistical properties. By modelling the signal and foregrounds simultaneously, this framework naturally accounts for their correlations and uncertainties, enabling robust separation of the faint 21-cm signal from dominant foreground contamination while propagating errors consistently through the inference process. Such a probabilistic approach is particularly powerful in scenarios with limited data or complex noise structures, where traditional sequential subtraction methods may introduce biases or underestimate uncertainties.

In contrast to blind methods, Bayesian approaches treat foreground mitigation as a probabilistic inference problem. The data are explicitly modelled as a combination of astrophysical foregrounds, instrumental contaminants, noise, and the cosmological 21-cm signal, with each component described by a likelihood and/or prior distribution. This allows one to quantify uncertainties on the recovered signal and to incorporate physical knowledge, such as the expected spectral smoothness of synchrotron emission or the frequency coherence of mode-mixing contaminants.

A widely used framework is Gaussian Process Regression (GPR)~\citep{Mertens2018}. In this approach, each component of the observed data is modelled as a Gaussian Process with its own covariance kernel: smooth kernels for Galactic and extragalactic foregrounds, intermediate coherence kernels for instrumental mode-mixing, and shorter coherence kernels for the 21-cm signal. Implemented in the LOFAR-EoR analysis pipeline, GPR has demonstrated the ability to recover power spectra close to the thermal noise limit, and to outperform blind methods such as GMCA in the presence of complex contaminants.

In addition to GPR-based approaches, a complementary class of emerging Bayesian frameworks are being developed which directly model the interferometric visibilities rather than post-processed images or power spectra. A key motivation for these approaches is that the correctness of any cosmological model depends on the fidelity of the instrumental calibration: inaccurate calibration alters the effective likelihood of the data and can bias downstream inferences. End-to-end Bayesian pipelines address this by linking calibration and inference within a single probabilistic framework, allowing instrumental, foreground, and cosmological parameters to be solved for jointly and self-consistently.  Two such pipelines that are actively being developed are BayesCal/BayesEoR and Hydra.

The BayesEoR\footnote{\url{https://bayeseor.readthedocs.io/en/latest/}} package \citep{2023MNRAS.520.4443B, 2024JOSS....9.6667S} implements such a fully forward-modelled, GPU-accelerated Bayesian framework that jointly infers the 21-cm power spectrum and astrophysical foreground parameters. By explicitly accounting for the covariance between the cosmological signal and the foregrounds induced by the instrument, BayesEoR enables statistically optimal recovery of the 21-cm signal without reliance on foreground avoidance \citep{2016MNRAS.462.3069S, 2019MNRAS.484.4152S}. It has been shown to recover unbiased power spectra across the accessible $k$-range in realistic HERA-like simulations and provides a natural means for model comparison through Bayesian evidence evaluation \citep{2019MNRAS.488.2904S, 2020MNRAS.492...22S}.

Related developments such as BayesCal \citep{2022MNRAS.517..910S,2022MNRAS.517..935S} apply the same probabilistic philosophy to calibration, marginalising over unmodelled sky emission during the calibration process itself. This approach suppresses calibration-induced spectral systematics by several orders of magnitude relative to standard methods and yields statistically consistent gain solutions. Together, these frameworks demonstrate the potential of end-to-end Bayesian inference for achieving the calibration accuracy and statistical robustness required for SKA-Low 21-cm cosmology. While still under active development, they represent one of several promising directions for future SKA analysis pipelines, where maintaining a diversity of complementary methodologies and models will be essential for achieving validated, cross-checked results \citep[e.g.][]{2025MNRAS.541.2262S} and ensuring cross-pipeline consistency within SKA data analysis workflows \citep[e.g.][]{2016ApJ...825..114J}.

Hydra\footnote{\url{https://hydraradio.github.io/Hydra/}} is a Bayesian inference framework for 21-cm cosmology based on the Gibbs sampling technique. This allows statistical samples to be drawn from complicated high-dimensional posterior distributions by carving them up into a set of simpler, lower-dimensional conditional distributions, each of which can be sampled from efficiently. The sampler then iterates over these distributions, building up a set of samples from the full joint posterior distribution as it does so. The main benefit of Gibbs sampling is that a detailed, physics-based forward model of the data can be specified, and full statistical uncertainties on all of the parameters propagated without approximation. Hydra implements the various conditional distributions needed to build a complete Gibbs sampler for 21-cm visibility data (as well as autocorrelation data, e.g. from MeerKAT and SKAO-MID), ranging from foreground and point source parameters \citep{2024MNRAS.535..793B, 2024RASTI...3..607G, 2025MNRAS.544.2419N}; calibration and reflection systematics parameters \citep{2024MNRAS.534.2653M, 10.1093/rasti/rzag024}; primary beam models \citep{2024RASTI...3..607G, 2025RASTI...4...42W}; to the 21-cm power spectrum \citep{2023ApJS..266...23K}. It has so far been demonstrated on simulated data for inference problems with hundreds of thousands to millions of parameters. A self-contained implementation for delay power spectrum analyses, of the kind used by HERA, is available as {\tt hydra-pspec}, and can be used to disentangle foregrounds, the 21-cm signal, and reflection systematics on a per-baseline basis \citep{2023ApJS..266...23K, 2024MNRAS.535..793B}. An additional benefit is that ``in-painting'' of flagged data, separation of foregrounds, filtering of systematics, and complete covariance matrix/window function estimates for the delay power spectrum are natural byproducts of the sampling process, meaning that several steps of `traditional' power spectrum estimation pipelines can be replaced with this single tool.

\subsubsection{Machine learning based approaches}\label{sec:signalseparation_ML}
Several machine learning approaches, particularly those based on convolutional neural networks (CNNs), have been developed to separate the faint 21-cm signal from overwhelming foreground contamination. Architectures such as autoencoders and the U-Net are trained on simulated datasets to distinguish the complex spatial and frequency patterns of the cosmological signal from the foregrounds. For example, convolutional denoising autoencoders can hierarchically learn sophisticated features to overcome instrumental effects and accurately separate the signal \citep{li2019separating}. More commonly, U-Net models are used to recover cosmological information from data that has already been partially cleaned by traditional techniques like PCA \citep{makinen2021deep21}. Other approaches have introduced novel preprocessing steps, such as using the temperature difference between adjacent frequency channels to reduce the dynamic range of the foregrounds before feeding the data into a U-Net \citep{shi202421}. To better prepare these networks for real-world application, some studies train their models on simulations that incorporate data-driven systematic effects, such as foreground residuals derived from actual LOFAR observations \citep{gao2025extracting}. In addition to direct separation, machine learning has also been used to improve conventional methods. For example, autoencoders can learn data-driven covariance models to improve GPR-based mitigation frameworks \citep{mertens2024retrieving,acharya202421}. These varied approaches have demonstrated a robust capability to recover the 21-cm signal, often outperforming purely conventional techniques.

More advanced frameworks aim to perform a full field-level reconstruction of the 21-cm signal map, including the modes within the foreground wedge that are typically lost. The \texttt{SERENEt} framework \citep{bianco2025deep}, for example, uses a dual-network approach: it first identifies neutral and ionised regions \citep[see e.g.,][]{giri2018optimal,giri2025mapping,bianco2024deep} and then uses this segmentation map as a prior to enhance the final signal recovery and reconstruct the full map. Similarly, other U-Net based methods have been developed to reconstruct signal modes that were completely removed in a foreground avoidance strategy, aiming to recover the morphology of ionised bubbles \citep{gagnon2021recovering}. These map-level recovery techniques are crucial for enabling imaging studies with current and next-generation instruments. These advanced reconstruction methods fundamentally leverage the non-Gaussian nature of the 21-cm signal, where mode-coupling allows information from observed scales to inform the reconstruction of unobserved, foreground-dominated scales.

Beyond direct map recovery, other techniques combine machine learning with physical models for even more sophisticated inference. One such approach utilises deep generative models, such as variational diffusion models, which are trained on simulations to learn the mapping from foreground-contaminated observations to the true underlying 21-cm field \citep{chen2025field}. An alternative strategy uses a gradient-based sampler within an Effective Field Theory (EFT) framework to jointly infer both the initial cosmic density field and astrophysical bias parameters directly from the observed data \citep{qin2025effective}. 
In contrast, comparatively simple Artificial Neural Networks (ANNs) have been employed for parameter inference, wherein the global 21-cm signal is extracted from foreground-contaminated data to estimate reionization parameters \citep{choudhury2020extracting, choudhury2021using, Tripathi2024, Tripathi2024_Samp}. This approach has subsequently been extended to recover signal parameters from mock 21-cm power spectra incorporating the effects of foreground contamination \citep{2022Choudhury}. For a broader discussion of machine learning applications for the SKA, we refer the interested reader to the \citet{Acharya02.2026.SKA}.

\subsubsection{Hybrid approaches}\label{sec:hybrid_approach}
Hybrid approaches often integrate "foreground avoidance" and "foreground subtraction" \citep{2018Kerrigan}. Foreground avoidance filters out contaminated modes in Fourier space, particularly the "foreground wedge," but this can also lead to a loss of cosmological information. Foreground subtraction, conversely, relies on explicit modelling of foregrounds, the accuracy of which is critical for effective removal \citep{2018Kerrigan}.

Advanced hybrid techniques include the Hybrid Foreground Residual Subtraction (HyFoReS), which targets beam-induced foreground residuals by cross-correlating linearly filtered data \citep{2018Kerrigan}. Statistical and machine learning methods are also widely used. Principal Component Analysis (PCA) and Singular Value Decomposition (SVD) are common blind component separation techniques, though their ability to achieve complete separation of foregrounds and the 21-cm signal is often limited. To enhance this, the "Singular Vector Projection (SVP)" method, a semi-blind PCA-based approach, incorporates a priori information \citep{2023Zuo}. Similarly, Bayesian semi-blind methods like HI Expectation-Maximization Independent Component Analysis (HIEMICA) extend techniques from Cosmic Microwave Background (CMB) maps to 3D 21-cm cosmological signals, providing robust Bayesian inference of power spectra \citep{2016Zhang}.

\subsubsection{Foreground separation from the MAPS}\label{sec:fg_maps}
In the MAPS-space, foregrounds have a distinctly different spectral behaviour than the EoR 21-cm signal. The signal is expected to be predominantly \textit{localized} within a typical $\Delta \nu $ range of  $ 0 - 2 \, \rm{MHz} = [\Delta\nu]_{\rm corr}$, and its amplitude drops substantially and is close to zero at large $\Delta \nu$. On the other hand, the foregrounds generally vary smoothly and remain correlated over a large $\Delta \nu$ range. The key idea is to model the foregrounds from the measured $C_\ell(\Delta\nu)$ in the range $\Delta \nu > [\Delta\nu]_{\rm corr}$, and using the best fit model to predict the foregrounds in the range $\Delta \nu \le [\Delta\nu]_{\rm corr}$. One can then subtract the foreground prediction and place constraints on the 21-cm signal. The error in foreground prediction can be easily predicted and propagated. The key benefit lies in the fact that the foregrounds are modeled from a region where the 21-cm signal is negligible, and therefore, it leads to a very small signal loss. One is free to choose a model for the foreground fitting and prediction, for example, \cite{Elahi2023b} chose a polynomial fitting and a GPR for this, both yielding very similar results.

One limitation of this method is that the decorrelation length scale $[\Delta\nu]_{\rm corr}$ is considerably larger for large angular scales (small $\ell$ or, equivalently small $k_\perp$). For a typical MWA Phase II configuration, EoR simulations (e.g. \citealt{Mondal2017}) shows $[\Delta\nu]_{\rm corr} \approx 4\, {\rm MHz}$ at the smallest $\ell$. Predicting foregrounds over such large frequency range $0-4\, {\rm MHz}$ has large uncertainties and can potentially lead to systematics from mismodelling. That being said, the method is promising for the smaller scales at which the EoR 21-cm signal is localized. One must note that modelling foreground from $C_\ell(\Delta\nu)$ at small scales is difficult due to the instrument's frequency dependence, which creates oscillatory patterns, and tapering the primary beam, as done in TGE, or other machine-learning-based modelling and prediction can be effective in such cases. 

\subsubsection{Signal loss and uncertainty quantification}\label{sec:signal_loss}
Foreground subtraction methods inherently face two fundamental risks: (i) the inadvertent removal of part of the 21-cm signal together with the contaminants (signal loss) and (ii) the incomplete subtraction of foreground components, leaving behind residual contamination. Both effects can introduce systematic biases into the recovered power spectrum, potentially leading to an underestimation of the signal or even false detections. Therefore, the success of a foreground mitigation strategy should be evaluated in terms of how effectively it suppresses both signal loss and residuals and how reliably the associated uncertainties are quantified.
Signal injection tests, in which simulated 21-cm signals are added to real data and passed through the analysis pipeline to assess recovery, are increasingly used to evaluate signal loss in 21-cm studies. For example, \cite{2021MNRAS.500.2264H} applied this approach in LOFAR analyses, comparing blind methods such as FastICA and GMCA with the Bayesian GPR. Similar strategies are beginning to emerge in other datasets, indicating that signal injection tests are becoming a valuable tool for validating foreground mitigation analysis.
\subsection{Interplay of foregrounds and the missing channels}
\label{sec:fg_missch}
Missing frequency channels pose a serious problem for in power spectrum estimation. In `delay space analysis', the measured visibilities are Fourier transformed along frequency to estimate the delay space visibilities, which are then used to estimate the power spectrum \citep{2004ApJ...615....7M, parsons2012per}. In this approach, the missing frequency channels introduce ringing artifacts in the delay space visibilities and corrupt the estimated power spectrum. Several algorithms are used to deal with this challenge \citep{Parsons2009, trott2016, 2019PASA...36...26B, 2021MNRAS.500.5195E, 2023ApJS..266...23K}

The missing channels do not pose a problem in the visibility-correlation approach adopted in TGE, where one first correlates the visibilities in the frequency domain to estimate the MAPS $C_\ell(\Delta\nu)$, and then perform a Fourier transform along $\Delta\nu$ to estimate the 21-cm power spectrum $P(k_\perp,k_\parallel)$ (Section~\ref{sec:viscorr}). Even if there are many frequency channels $\nu$ missing, the estimated $C_\ell(\Delta\nu)$ does not have a missing frequency separation $\Delta\nu$. Therefore, in this approach, the power spectrum does not show artifacts due to the missing channels. The key idea is that it is not essential to compensate for the missing frequency channels, as the power spectrum can be estimated using only the available channels. \cite{2019MNRAS.483.5694B} have used simulations to show it is possible to estimate $P(k_\perp, k_\parallel)$ without any artifacts even when $80 \%$ of randomly chosen frequency channels are flagged. It also performs well on actual data \citep{Pal2020, Pal2022, Elahi2023, Elahi2023b, Elahi2024}.

MWA, the SKA-Low precursor, has a periodic pattern of flagged channels, for which the delay space analysis introduces horizontal streaks in  $P(k_\perp, k_\parallel)$  \citep{Paul2016, Li2019, Trott2020, Patwa2021}. \cite{Chatterjee2024} used simulated data, keeping exactly the same flagging as the actual MWA data, to show that TGE can recover $P(k_\perp, k_\parallel)$ without any artifacts. However, for the actual data, they also find a pattern of spikes along $k_{\parallel}$, the period of which corresponds to the period of the pattern of flagged channels. Although the amplitude of these artifacts is much smaller compared to those in the delay-space analysis, it suffices to contaminate a large region of the EoR window. \cite{Elahi25_missingchan} demonstrated that these tiny artifacts arise from a combination of two factors, namely the periodic pattern of flagged channels and the strong spectral dependence of the foregrounds. They suggest that it is possible to mitigate the artifacts in the power spectrum if we can reduce the overall amplitude and spectral variation of the foreground contamination in the data. They propose the technique named `Smooth Component filtering (SCF)' where the idea is to simply model the foregrounds as a smooth component of the data and subtract it from the visibility data. In the actual implementation, they have convolved the visibilities along the frequency axis with a Hann window -- with a data-driven choice of a 2~MHz smoothing scale -- to obtain the smooth component of the data. Then they subtract the smooth component from the visibility and use the residual visibilities to obtain the MAPS $C_\ell(\Delta\nu)$ and from it the power spectrum $P(k_\perp, k_\parallel)$. A remarkably clean, \textit{noise-dominated}, EoR window is found with this technique. A caveat of this approach is that one loses all signal smoother than the chosen smoothing scale, and it is necessary to \textit{avoid} the low $k_\parallel$ region to place constraints on the 21-cm signal. An alternate, and desiarable approach would be to accurately model the foregrounds from an image and subtract these from the visibility data. Such an approach typically requires observations with a high angular resolution and sensitivity, something that will be possible with the SKA-Low. 

\vspace{-8pt}
\section{Mitigation strategies for SKA observation}\label{sdc3a}
The section will provide a review of results from the recent SDC3a challenge, current challenges and limitations in foreground subtraction, and an outlook on future directions.

\subsection{Review of the SKA Data Challenges 3a results}\label{sdc_review}

\subsubsection{Definition of the challenge}
The Square Kilometre Array Science Data Challenge 3a (SDC3a)\footnote{\url{https://sdc3.skao.int/challenges/foregrounds}}\citep{2025SDC3a} was organised by the SKAO to assess the community’s ability to recover the faint cosmological 21-cm signal from highly contaminated low-frequency interferometric data, representative of upcoming SKA-Low EoR observations.

Participants received realistic end-to-end simulated SKA-Low visibilities spanning $106$--$196$~MHz, containing the 21-cm signal, diffuse Galactic synchrotron and free--free emission, extragalactic foregrounds (in-field and out-of-field sources), thermal noise, and calibration errors. The task was to reconstruct the cylindrically averaged 21-cm power spectrum, $P(k_{\perp}, k_{\parallel})$, and quantify uncertainties and biases. Teams submitted six cleaned, noise-subtracted 2D power spectra over non-overlapping 15~MHz sub-bands across the 90~MHz range, using only the central $4 \times 4$~deg of the $8 \times 8$~deg field to reduce noise. Error bars (1$\sigma$) were assumed Gaussian and uncorrelated, with $k_{\parallel}$ and $k_{\perp}$ binning specified by the organisers.
 
\subsubsection{Analysis of the challenge results}
The SDC3a was a community-wide challenge organised by the SKAO to evaluate how effectively different foreground-removal techniques can recover the faint 21~cm EoR signal in the presence of bright astrophysical foregrounds and instrumental noise. Seventeen teams participated, employing diverse methodologies.

\subsubsection*{Methodologies Employed}
Teams adopted a variety of strategies for signal recovery. Some performed visibility-domain cleaning by subtracting point-source or continuum models prior to imaging, while others used image-domain approaches such as GPR, PCA, U-Net or Transformer networks, polynomial fitting, or semi-supervised beam removal. Several groups focused on foreground avoidance by analysing only regions of $k$-space less contaminated by foregrounds, whereas a few employed forward-modelling frameworks to jointly model the EoR signal and foregrounds.

\subsubsection*{Performance Metrics}
Overall performance varied across submissions. Seven teams achieved median errors below the true EoR power spectrum, indicating effective foreground mitigation. The most accurate recovery yielded a residual power spectrum of $4.2_{-4.2}^{+20} \times 10^{-4}\,\rm{K}^2\,h^{-3}\,\rm{cMpc}^3$. Although most teams reported uncertainties, these were generally underestimated, with the best estimates capturing only about $60 \pm 20\%$ of the true errors.

\subsubsection*{Insights and Implications}
Cross-comparison revealed residual biases and highlighted the need for improved uncertainty estimation. The diversity of approaches suggests that the community is broadly prepared for SKA-Low EoR data, though accuracy still varies between techniques. Machine-learning models, including GPR, U-Nets, Transformers, along with forward-modelling approaches, showed strong promise for robust foreground mitigation. Continued advances in both foreground-removal methods and uncertainty quantification will be essential for reliable EoR science.

In summary, SDC3a demonstrates that current pipelines can recover the cosmological 21~cm signal from realistic SKA-like data, but further refinement in calibration modelling, error propagation, and systematic control is required to achieve high-fidelity EoR measurements with SKA-Low.

\subsection{Current Challenges and Limitations}\label{sec:limitation}
\subsubsection{Limitations of current foreground models}
Section~\ref{sec_mitigation} of this chapter discusses various mitigation strategies for modelling foregrounds from targeted or survey mode observations with the SKA. It also discusses various instrumental and observational biases that may make it challenging to accurately model the spectral and spatial structure of the foreground, as well as how different sky source subtraction approaches address these obstacles. Therefore, foreground separation in SKA observations is tightly coupled with precision calibration, as both rely on accurately modelling the spectral and spatial structure of the low-frequency sky.

However, instrumental systematic effects corrupt the spectrally correlated nature of foregrounds. The primary beam of dipole antenna arrays exhibits intrinsically unstable element gains that vary with time, frequency, and direction, making accurate beam modelling non-trivial. While mutual coupling and faulty elements introduce complex and unstable beam patterns. These effects distort the assumed spectral smoothness of bright off-axis sources, complicating foreground subtraction \citep{Thyagarajan2016, Joseph2020, Nasirudin2020, Trott2020, Fagnoni2021, 2025MNRAS.541.3993B, Chokshi2024}. Additional systematics, such as cable reflections and cross-coupling,  introduce frequency-dependent ripples that can mimic or contaminate the cosmological signal \citep{Kern2020}. Although hardware upgrades have reduced these effects, accurate and physically informed beam and instrument models are still required for precise calibration \citep{2024MNRAS.534.2653M, Rath2025}.

\subsubsection{Residual error assessment}
Mitigating and quantifying residual systematic errors remains crucial, as they introduce excess noise that limits the detection of the 21-cm signal. These residuals arise from numerous factors, including gain instability, polarised foreground leakage, ionospheric distortions, and instrumental imperfections. It is essential to understand and quantify their impact in order to isolate the cosmological signal from contamination. Over the past decade, precursor and pathfinder instruments around the world have made significant progress in understanding and quantifying this excess noise through detailed simulations and experimental analysis. Comprehensive end-to-end simulations for SKA-Low demonstrate that calibration inaccuracies exceeding $0.1\%$  lead to significant foreground leakage into the EoR window \citep{Mazumder2022}. While realistic ionospheric modelling further shows that the median ionospheric offset must remain below $\theta_{\rm MIO} \leq 0.1''$ for unbiased power spectrum estimation \citep{Pal2025}. These finding highlights the critical need for further development of calibration algorithms for future SKA1-Low observations. Even minute instrumental imperfections, such as $0.1\%$ cable length errors or $5\%$  element failures, can introduce $\sim 10\%$ contamination across k-modes \citep{Nasirudin2022, Oscar2024}. Furthermore, \citet{Oscar2025} illustrated that the necessity for station beam models to achieve beam model precision at the level of four to five significant digits is vital for effective foreground removal.

\subsection{Future Directions in Foreground Mitigation}\label{sec:future_scope}
Physics-informed machine learning (PIML) is an emerging and powerful approach for 21 cm foreground removal that blends physical modeling with data-driven techniques to recover the faint cosmological signal from bright foregrounds. PIML methods embed physical insights such as the spectral smoothness of foregrounds and statistical properties of the 21 cm signal directly into neural network architectures or loss functions. For example, Physics Informed Neural Networks (PINNs) incorporating radiative transfer physics \citep{2023Korber} or signal morphology constraints enable models to generalize across different instrumental conditions and simulation setups.

Simulation-based inference (SBI) has also become a key framework for 21 cm analysis, particularly in scenarios where complex noise, instrumental systematics, and foregrounds make traditional likelihood-based methods impractical \citep{2024Saxena}. SBI uses forward simulations to generate mock observations spanning cosmological and instrumental parameter spaces and trains neural density estimators to approximate posterior distributions without requiring analytic likelihoods. This flexibility makes SBI well suited for modeling the non-Gaussian data typical in 21 cm cosmology \citep{2023Prelogovi}.

In global 21 cm studies, SBI frameworks have been applied to jointly infer signal and foreground parameters using simulated sky spectra that include realistic astrophysical foregrounds, ionospheric effects, and instrumental distortions \citep{choudhury2020extracting, choudhury2021using, Tripathi2024, Tripathi2024_Samp, 2026Tripathi}. These neural network-based models accurately recover astrophysical parameters and efficiently explore high-dimensional parameter spaces. Similar methodologies have also been extended to parameter recovery from mock 21-cm power spectra affected by foreground contamination \citep{2022Choudhury}.

In recent applications, such as REACH and similar 21-cm experiments, SBI has been used to jointly infer foreground and signal parameters, achieving constraints comparable or superior to classical approaches while requiring significantly fewer simulations. By incorporating physically motivated models and leveraging deep-learning-based posterior estimation, SBI improves the robustness of foreground mitigation pipelines and offers a scalable pathway to reliable signal recovery across multi-instrument and multi-epoch datasets in upcoming 21-cm surveys \citep{2024Saxena}.

\section{Conclusion}\label{conclusion}
The SKA-Low, with unparalleled sensitivity and resolution, will enable accurate modeling of complex foreground structures. This chapter discusses the primary challenge to modeling and current mitigation strategies for foreground subtractions. This mitigation strategy combines wedge avoidance, precision subtraction via direction‑dependent calibration and beam correction, and advanced statistical separation to control leakage beyond the EoR window reviewed in this chapter. The lessons learned from pathfinder observations reviewed in this chapter around the world have pushed the boundaries of understanding and quantifying residual systematic errors. The global community has made significant progress in developing data analysis pipelines capable of subtracting foregrounds from realistic SKA-like observations. Realistic end-to-end simulation tools, including instrumental and propagation errors, are crucial to testing and validating data analysis pipelines. The development of physics-informed machine learning and simulation-based Bayesian inference represents a paradigm shift, enabling the joint modeling of cosmology, foregrounds, and instrumental effects. By combining physical insights with advanced statistical and computational tools, these next-generation methods promise precise extraction of the faint 21-cm signal, advancing our understanding of the early universe.


\section{Author contributions}
A. Datta, A. Chakraborty, A. Tripathi, S.K. Pal, and F. Mertens contributed to shaping the structure, organized the contributions, and edited the content. We give the contributions of the authors to every section below.
\paragraph{Sec~\ref{sec:intro}} A. Tripathi, S.K. Pal and A. Datta wrote this section.

\paragraph{Sec~\ref{sec:foreground_sources}} R. Sagar, A. Chakraborty and S. Choudhuri contributed to Sec~\ref{sec:gal_total}(Galactic emission in total intensity). V. Jelić wrote Sec~\ref{sec:polgal}. Z. Zhu and H. Shan wrote Sec~\ref{sec:extra-galactic}. S.K. Pal and A. Datta wrote Sec~\ref{sec:ionosphere}. E. Tolley contributed to the first two paragraphs of Sec~\ref{sec:rfi} and R. Sagar contributed to the third paragraph at Sec~\ref{sec:rfi}. Y. Mao, Z. Zhu and H. Shan wrote Sec~\ref{sec:fcode}. 

\paragraph{Sec~\ref{sec_mitigation}} R. Nunhokee, S. Munshi, and A. Chakraborty contributed to Sec~\ref{sec:compare_tech}\,(Avoidance methods). K. M. A. Elahi wrote Sec~\ref{sec:viscorr} (Visibility-correlation approach). A. R. Offringa, R. Nunhokee and E. Ceccotti wrote Sec~\ref{sec:source_subtraction} and R. Sagar writing the Direction-dependent calibration subsection (Sec~\ref{dd-calibration}). F. Mertens, S. K. Giri, J. Burba, T. Ito, Y. Mao, Z. Zhu and H. Shan contributed to Sec~\ref{sec:sigsep} (Signal separation techniques). K. M. A. Elahi wrote subsection~\ref{sec:fg_maps}. L. Zhang contributed to subsection~\ref{sec:bays_approach}, with J. Burba and P. H. Sims. S. K. Giri and A. Tripathi contributed to subsection~\ref{sec:signalseparation_ML}. K. M. A. Elahi wrote Sec~\ref{sec:fg_missch} (Interplay of foregrounds and the missing channels).
\paragraph{Sec~\ref{sdc3a}} A. Tripathi contributed to the Sec~\ref{sdc_review}\,(Review of the SKA Data Challenges 3a results) and Sec~\ref{sec:future_scope} (Future Directions in Foreground Mitigation). S.K. Pal wrote Sec~\ref{sec:limitation}\,(Current Challenges and Limitations).
\paragraph{Sec~\ref{conclusion}} S.K. Pal wrote this section with contributions from A. Datta and A. Tripathi.

We also thank the anonymous reviewer for their constructive feedback.


\bibliographystyle{abbrvnat}
\bibliography{chapter} 

\end{document}